\documentstyle[12pt,aaspp4,epsfig]{article}

\def\deg{\ifmmode^\circ\else$^\circ$\fi}

\def\mic{~$\mu$m}

\def\h0{H$_0$}
\def\q0{q$_0$}

\def\arcs{\ifmmode {''}\else $''$\fi}
\def\arcm{\ifmmode {'}\else $'$\fi}
\def\parcs{\sa=.07em \sb=.03em
     \ifmmode $\rlap{.}$^{\scriptscriptstyle\prime\kern -\sb\prime}$\kern -\sa$
     \else \rlap{.}$^{\scriptscriptstyle\prime\kern -\sb\prime}$\kern -\sa\fi}
\def\parcm{\sa=.08em \sb=.03em
     \ifmmode $\rlap{.}\kern\sa$^{\scriptscriptstyle\prime}$\kern-\sb$
     \else \rlap{.}\kern\sa$^{\scriptscriptstyle\prime}$\kern-\sb\fi}

\def\ebmv{E(B-V)}
\def\Msun{M$_{\odot}$}
\def\Myr{\Msun/yr}
\def\kp{{\rm K}$^{\prime}$}

\def\han {\mbox{{\rm H}$\alpha$}}
\def\hb {\mbox{{\rm H}$\beta$}}
\def\hgamma {\mbox{{\rm H}$\gamma$}}
\def\oiii{[O~{\sc iii}]}
\def\oii{[O~{\sc ii}]}
\def\oi{[O~{\sc i}]}
\def\nii{[N~{\sc ii}]}
\def\ha{\han}

\def\spose#1{\hbox to 0pt{#1\hss}}
\def\simlt{\mathrel{\spose{\lower 3pt\hbox{$\mathchar"218$}}
     \raise 2.0pt\hbox{$\mathchar"13C$}}}
\def\simgt{\mathrel{\spose{\lower 3pt\hbox{$\mathchar"218$}}
     \raise 2.0pt\hbox{$\mathchar"13E$}}}
\def\lsim{\rlap{$<$}{\lower 1.0ex\hbox{$\sim$}}}
\def\gsim{\rlap{$>$}{\lower 1.0ex\hbox{$\sim$}}}

\begin{document}

%\slugcomment{Draft Version 0.0:by \today}
%\pagestyle{myheadings}
%\markboth{DRAFT version 0.0: \today}{DRAFT version 0.0: \today}

\title{The Rest-Frame Optical Spectrum of MS 1512-cB58 
\footnote{
Data presented herein were obtained at the W.M. Keck Observatory, 
which is operated
as a scientific partnership among the California Institute of Technology, 
the University of California and the National Aeronautics and Space 
Administration.  The Observatory was made possible by the
generous financial support of the W.M. Keck Foundation.
}}

\author{Harry I. Teplitz\altaffilmark{2,}\altaffilmark{3},\\
Ian S. McLean\altaffilmark{4}, 
E. E. Becklin\altaffilmark{4}, 
Donald F. Figer\altaffilmark{5},
Andrea M. Gilbert\altaffilmark{6}, 
James R. Graham\altaffilmark{6},
James E. Larkin\altaffilmark{4}, 
N. A. Levenson\altaffilmark{7}, 
Mavourneen K. Wilcox\altaffilmark{4}}
\altaffiltext{2}{Laboratory for Astronomy and Solar Physics, Code 681, Goddard
Space Flight Center, Greenbelt MD 20771}
\altaffiltext{3}{NOAO Research Associate}
\altaffiltext{4}{Department of Physics and Astronomy, 
University of California, 
                 Los Angeles, CA, 90095-1562 }
\altaffiltext{5}{Space Telescope Science Institute, 
                  3700 San Martin Dr., Baltimore, MD 21218 }
\altaffiltext{6}{
Department of Astronomy,  
University of California, Berkeley,
601 Campbell Hall,
                 Berkeley, CA, 94720-3411}
\altaffiltext{7}{
Department of Physics and Astronomy,  
Johns Hopkins University,
                 Baltimore, MD  21218}

\begin{abstract}

  Moderate resolution, near-IR spectroscopy of MS1512-cB58 is
  presented, obtained during commissioning of the the Near IR
  Spectrometer (NIRSPEC) on the Keck II telescope.  The strong lensing
  of this $z=2.72$~galaxy by the foreground cluster MS1512+36 makes it
  the best candidate for detailed study of the rest-frame optical
  properties of Lyman Break Galaxies.  In eighty minutes of on-source
  integration, we have detected \ha, \nii $\lambda$6583,6548\AA, 
  \oi$\lambda$6300\AA, He I $\lambda$5876\AA, \oiii
  $\lambda$5007,4959\AA, \hb, \hgamma, \oii$\lambda$3727, and a strong
  continuum signal in the range 1.29-2.46\mic.
  
  A redshift of $z=2.7290\pm 0.0007$~is inferred from the emission
  lines, in contrast to the $z=2.7233$~calculated from UV observations
  of interstellar absorption lines.  Using the Balmer line ratios, we
  find an extinction of \ebmv=0.27.  Using the line strengths, we
  infer an SFR$=620\pm 18$~\Myr (\h0=75, \q0=0.1, $\Lambda =0$), a
  factor of 2 higher than that measured from narrow-band imaging
  observations of the galaxy, but a factor of almost 4 lower than the
  SFR inferred from the UV continuum luminosity.  The width of the
  Balmer lines yields a mass of $M_{vir}=1.2\times 10^{10}$~\Msun.  We
  find that the oxygen abundance is 1/3 solar, in good agreement with
  other estimates of the metallicity.  However, we infer a high
  nitrogen abundance, which may argue for the presence of an older
  stellar population.
 
\end{abstract}

\keywords{
cosmology: observations ---
galaxies: evolution ---
galaxies: individual:  MS1512-cB58
}

\section{Introduction}

The largest sample of high redshift galaxies was selected from
observations of the extinction of rest-frame far-ultraviolet light by
intrinsic and intergalactic absorption.  These Lyman Break galaxies
(hereafter LBGs; Steidel et al. 1996) may, in principle, be the
tracers of the global star formation history of the universe (Madau et
al. 1998).  However, the unquantified effects of dust extinction
present an obstacle to such interpretation.  For example, corrections
to the star formation rate (SFR) of LBGs have been suggested to be as
large as factors of 2--10 (Pettini et al.  1998, hereafter P98; Trager
et al.  1997).

In this paper we present the first observations of an LBG
with NIRSPEC, the near infrared spectrometer on the Keck II
telescope.  
We will use the rest-frame optical spectrum to obtain 
a model independent measure
of dust extinction, an estimate of the virial mass,
 and the star formation rate.
Spectra such as this one also provide the best method for
determining the metal abundance in LBGs.  We will examine 
the evolutionary status of cB58, and its implications 
as a ``typical'' LBG.  

We have chosen the brightest LBG as our first target (MS1512-cB58; see
Yee et al. 1996).  MS1512-cB58 is a gravitationally lensed starburst
galaxy at z=2.72.  It was discovered serendipitously, within the field
of the z=0.37 galaxy cluster MS1512+36, in spectra obtained for the
CNOC survey (Yee, Ellingson, \& Carlberg 1996).  MS1512-cB58 is
clearly extended with the morphology of a lensed arc (Williams \&
Lewis, 1997; Seitz et al., 1998).  It has an ultraviolet spectrum
representative of LBGs but due to a factor of $\sim 30$~magnification
(Seitz et al.  1998) it is orders of magnitude brighter than any other
object of its kind (V=20.6, \kp=17.8; Ellingson et al.  1996,
hereafter E96).  Like most LBGs, its UV spectrum (devoid of strong,
high ionization emission lines) rules out any non-stellar (AGN)
contribution to its flux.  Its apparent (lensed) star-formation rate
is 2417 \Myr~(\h0 = 75 km s$^{-1}$Mpc$^{-1}$, \q0 = 0.1; with
extinction correction) from the 1500\AA~continuum (Pettini et al.
2000, hereafter P2000).  Throughout this paper, unless otherwise
noted, a cosmology of (\h0 = 75 km s$^{-1}$Mpc$^{-1}$, \q0 = 0.1,
$\Lambda =0$) is assumed.

\section{Observations}

Near IR spectra were obtained using the NIRSPEC instrument on the 10-m
W. M. Keck II telescope  (McLean et al. 1998, 2000a).  
Spectra were obtained in the low resolution, long-slit mode, using a 
slitwidth of 3 pixels (0.57\arcs) at the (InSb) detector.
Observations were made in the H and K bands, as well as the wavelength
range between J and H (1.3-1.6\mic; hereafter N4 for the custom
NIRSPEC-4 filter) The seeing was 0.4\arcs, 0.4\arcs, 0.6\arcs~for N4, H, K
respectively.  MS1512-cB58 is highly elongated with a major axis $\sim
2$\arcs~long, but a minor axis only a few tenths of an arcsecond wide.
The slit was aligned with the long axis, while the unresolved 
(seeing broadened) width filled the slit.
H-band spectra were
obtained in 300 second exposures separated by $\sim 10$\arcs~nodding
along the slit; N4-band and K-band spectra were taken in 600 second
integrations.  Total integration times were 1200, 1800, and 1800 seconds 
in N4, H,and K respectively.

The data reduction procedure followed the steps outlined in McLean et
al. (2000b).  Wavelength calibration was good to 0.99\AA.  The output
pixel scale was 4.16\AA/pixel in H and K, and 2.08\AA/pix in N4.
Unresolved Argon arc-lamp lines had FWHM$=13.7$\AA, for a final
resolution of, for example, R$= 1300$~at 1.8\mic.  Nodded frames were
registered, and the final 1D spectrum was extracted using a Gaussian
optimal extraction filter.  Extracted spectra were divided by the
atmospheric absorption spectrum.  We obtained this spectrum from
similarly-reduced observation of a mostly featureless star (HR5569, an
A2V for the K-band; PPM130690, a G star for the H-band; and HR5630, an
F8V star for the N4-band), divided by a Kurucz (1993) model
atmosphere.  Emission-line equivalent widths (EW) were measured using
Gaussian fits to the line after subtraction of a local continuum.  The
flux of the line is then the EW multiplied by the continuum flux
density, which is well known from broad-band photometry (E96).  MS1512-cB58
is a homogenous (featureless) lensed image of the highest surface
brightness part of the source galaxy (Seitz et al. 1998); as such, our
measurements of the EW can scale directly to the photometry, since
light lost outside the slit is just a pure fraction of the total
light.  No real features fall outside the slit, 
as might be expected for an unlensed galaxy.

\section{Results}

Figures 1 - 3 show the complete 1.28-1.53\mic, 1.55--1.97\mic~and
2.05-2.46\mic~spectra of MS1512-cB58.
Just as LBG spectra in the rest-frame
UV are remarkably similar to nearby starbursts, so too the optical
spectrum of MS1512-cB58 is very similar to local irregulars or late type
spirals.  Table \ref{tab: fluxes}~lists the wavelengths and fluxes of
the measured emission lines.  

The mean redshift calculated from the emission lines is $2.7290\pm
0.0007$, in contrast to the $z_{abs}=2.7233\pm 0014$~reported in Yee
et al. (1996) from the rest-frame UV interstellar absorption lines.
P98 find that velocity shifts on the order of several hundred km
s$^{-1}$~between interstellar absorption lines and H~{\sc ii}~emission
lines are typical in LBGs.  P2000 confirm that this shift is seen in
MS1512-cB58 as well, finding $z_{em}=2.7296 \pm 0.0012$~in good
agreement with our result.  Since the measurement of the systemic
redshift of MS1512-cB58 has only recently been available, several
observations have been made in recent years assuming
$z_{em}=z_{abs}$~(e.g. Bechtold et al. 1997; Frayer et al.
1997;.Nakanishi et al.  1997).  In most cases the observations are
still valid despite the shift.

P2000 find evidence in the UV spectrum for intervening absorption
systems at z=0.829 and z=1.339.  The optical spectrum shows no
emission features associated with those systems.  
However, the absorbing galaxy might be several arcseconds
away from MS1512-cB58, and so might not have fallen in the slit.

\subsection{Dust Extinction and Virial Mass}

The extinction correction due to dust is crucial to understanding the
star formation activity occurring in the object. E96 infer \ebmv=0.3
from broad band photometry compared to synthetic galaxy spectra, and
the LMC extinction curve of Fitzpatrick (1986).  The best extinction
correction for MS1512-cB58 derived from the ultraviolet continuum
(P2000) yields \ebmv=0.24, assuming the same extinction law; or
\ebmv=0.29 for the Calzetti (1994) reddening law.  Both of these
measurements, however, are dependent on the slope of the model 
continuum.

We measure \ha:\hb=$3.23 \pm 0.4$~(note that the \hb~line is
measured with $SNR<10$).  It is assumed that the underlying stellar
absorption of the Balmer lines is small, and of equal magnitude for
both lines (Olofsson 1995).  Assuming Case B recombination for H I at
T=10000 K and $n_e$(cm$^{-3}$)$\le 100$, the intrinsic 
\ha:\hb~ratio is 2.86 (Osterbrock 1989), implying \ebmv $\simeq 0.27$~using
the Calzetti law (and very little change using the LMC law).  We
measure \hgamma:\hb $\sim 0.41 \pm 0.06$, compared to an expected
ratio of 0.45.

The strong emission lines allow us to measure the velocity dispersion.
Fitting Gaussian profiles to (e.g.) \ha~we can measure
$\sigma=$FWHM/2.355.  The width must then be corrected for the
instrumental resolution.  
We find $FWHM_{\mbox{Arc line}}=13.72$\AA~and 
$FWHM_{\mbox{\ha}}=20.8$\AA~(observed).  Thus, in the rest frame, 
$\sigma_{v}=81$ km s$^{-1}$.
Seitz et al. (1998) find a half-light radius $r_{1/2}=1.4
h_{75}^{-1}$kpc.  From these values, we obtain a virial mass
$M_{vir}=1.2\times 10^{10}$\Msun.  As expected, this value is in good
agreement with the masses of LBGs at $z\sim 3$ (P98).

\subsection{Chemical Abundance}

Kobulnicky et al.  (1999; hereafter KKP99) give a prescription for
determining chemical abundance estimates from the integrated spectra of high
redshift galaxies.  The best estimates are obtained for those spectra
that detect \oiii$\lambda 4363$, which directly measures the electron
temperature of the ionized gas.  We do not detect this line in our
spectrum, with a limiting flux of $0.31\times 10^{-16}$erg 
cm$^{-2}$s$^{-1}$~($3\sigma$).  Reasonable (0.2dex) estimates are
obtained in the absence of that line if the \oiii $\lambda 5007$,
\hb,and \oii $\lambda
3727$~lines can be measured, and we proceed with that limitation 
from the outset.

We calculate extinction corrected ratios of \oii /\hb$=3.52$~and 
\oiii$\lambda 5007$/\hb$= 3.57$.  We then calculate the quantity
$
R_{23}\equiv (I_{3727} + I_{4959} + I_{5007})/I_{H\beta}  
$.
$R_{23}$~has a direct relationship to the the oxygen abundance (Pagel
et al.  1979).  In the case of MS1512-cB58, $R_{23}=8.334$.  Using the
empirical calibration of Zaritsky et al. (1994), where $x\equiv
log(R_{23})=0.921$, 
\begin{equation}
12+log(O/H) = 9.265-0.33x-0.202x^2-0.207x^3-0.333x^4,
\end{equation}
or $12+log(O/H)\sim 8.39$.   However, the relationship between $R_{23}$~and oxygen
abundance can be double-valued in this low metallicity regime (McGaugh 1991).
So, following KKP99, we use the relationship between the 
\oiii$\lambda 5007$/\nii$\lambda 6584$=$10^{1.16}$~and the oxygen abundance 
 to confirm that the inferred oxygen abundance
is reasonable (Edmunds \& Pagel 1984).

The solar value of $12+log(O/H)$~is 8.89, and thus 
cB58 has a metallicity 0.32 solar, which is broadly consistent with the 
$Z/Z_{\odot}\sim 0.25$~found by P2000.  Frayer et al. (1997) find a similar
result, $Z/Z_{\odot}\sim 0.2 - 0.6$, based on the spectral energy 
distribution (SED) and extinction. 

The emission line ratios also yield the nitrogen abundance in MS1512-cB58.
The N/O ratio is typically assumed to be equal to the
N$^+$/O$^+$~ratio.  Using our estimates above, we can calculate N/O if
we assume an electron temperature that is consistent with the oxygen
abundance.  Using KKP99's figure 5, we find $T_e(O^{++})\simeq 9800$~K,
and then from
Vila-Costas \& Edmunds (1993; hereafter VCE), we infer
$T_e(O^+)=11400$~K from their Table 1, and
\begin{equation}
log(N^+/O^+) = \log \frac{F_{6548}+F_{6584}}{F_{OII}} + 0.307 - 0.02\log t_{OII} - \frac{0.726}{t_{OII}},
\end{equation}
where $t_{OII}=T_e(O^+)/10^4$.  Thus we find $log(N^+/O^+)=-1.24$.
This value of the nitrogen abundance is well below the solar value,
consistent with the metallicity inferred from the oxygen abundance.
However, the nitrogen abundance is still higher than might be
expected for a galaxy as young as MS1512-cB58 has been predicted to be.  The
oxygen:nitrogen abundance ratio is most consistent with local ($>
1$Gyr old) Scd galaxies (VCE), though within the observed range of
local, possibly younger, irregulars.  Further, the relatively high
nitrogen abundance suggests that an older population of stars is
necessary to produce it.  The ratio $log(N/O)>-1.4$~suggests the presence
of a secondary phase of nitrogen production, from intermediate mass stars
(Kobulnicky \& Zaritsky 1999).  
This result is in conflict with spectral synthesis modeling by
E96 which showed that the SED of MS1512-cB58 can be explained without
a population of stars older than 1 Gyr.  
However, E96 only rule out an older
population at the 2$\sigma$~confidence level, so we conclude 
the older population of stars is likely.

\subsection{Star Formation Rate}

P2000 infer an observed (without correcting for the 
lensing magnification factor of 30) 
SFR=2417 $h_{75}^{-2}$\Myr~(\q0 = 0.1) in MS1512-cB58 from the 
1500\AA~continuum.  This estimate included a factor of 7 correction
for dust extinction (from an assumed \ebmv=0.24).  

Kennicutt (1983) relates the SFR to \ha~luminosity, assuming a
Salpeter Initial Mass Function (IMF) with an upper mass cutoff of
100M$_{\odot}$, by: $\mbox{SFR(\Msun yr}^{-1}) =
L(\mbox{\rm{H}}\alpha) / 1.12\times 10^{41}\mbox{erg~s}^{-1} $ .
Kennicutt assumed 1.1 magnitude attenuation between \ha~and radio
fluxes, similar to that implied by our extinction measurement.  We
neglect stellar absorption of \ha, which is likely to be only
1-5\AA~EW.  We thus infer an (observed, lensing magnified) SFR=620$\pm
18$~$h_{75}^{-2}$ \Myr~(\q0=0.1).  This rate is a factor of 2.18
higher than that measured by Bechtold et al. (1997), using narrow-band
imaging.  This difference is the result of an atmospheric absorption
trough at the wavelength of redshifted \ha, that obscures
approximately half the light.  This absorption was not accounted for
by Bechtold et al., who assumed z=2.7233 rather than z=2.729 for
\ha~(fortunately, their filter was wide enough to still detect the
line).

While we measure a higher SFR from \ha~than Bechtold et al., we do
not resolve the question of the discrepancy with the rate inferred
from the UV continuum, a factor of 3.9 higher.  This discrepancy may
 be seen in the calculation of SFRs for local starbursts as well.
Meurer et al. (1995) find UV fluxes 10 times the value predicted by
Leitherer et al. (1995) given measured \ha; recall that
similar models of Leitherer et al. (1999; the updated
{\it Starburst99}) were also used to
calculate the SFR in MS1512-cB58.  Bechtold et al. speculate that a number of
unlikely possibilities could logically account for the difference.  We
find no compelling resolution to the problem in the optical spectrum.

For vigorously star-forming galaxies with moderate to solar
metallicity, the CO emission from the gas reservoir 
should be observable with existing ground based radio telescopes.
However, two attempts to
measure the CO(3-2) line in MS1512-cB58 have yielded only upper limits
(Frayer et al. 1997, Nakanishi et al. 1997).
The lack of detectable CO emission in MS1512-cB58 may be the result of either
a high CO-H$_2$~conversion factor, or highly efficient formation of
stars from the neutral hydrogen gas.  Frayer et al. suggest the star
formation efficiency (SFE) of MS1512-cB58 is a factor of 20 larger than in
local starbursts.  However, following the arguments in Frayer et al.,
if we take our measurement of metallicity, $Z/Z_{\odot}=0.32$, and of
SFR=620\Myr, we would find only a factor of 1.7 increase in the SFE.
Our assumptions would not address the high
$L(UV)/L^{\prime}(CO)$~ratio.  We can, however, bring the $L($\ha
$)/L^{\prime}(CO)$~ratio closer to the local values (following
Nakanishi et al.).

\section{Conclusions}

While it is difficult to draw general conclusions from observations of
one object, the rest-frame optical spectrum of MS1512-cB58 affords us
a preview of near-IR spectroscopy of LBGs.  As expected (Seitz et al.
1998, P2000), cB58 does not appear to be a ``proto-galaxy'', but a
galaxy with significant metals produced in at least two phases.  We
find that the properties of MS1512-cB58 are consistent with the
interpretation of LBGs as progenitors of modern-day elliptical
galaxies.  Previous studies have also shown that the mass, SFR, and
extinction of LBGs are appropriate in that context (i.e. P98).  We
have obtained the first measure of metal abundance in an LBG from the
promising $R_{23}$~method.  The metallicity of $1/3$~solar is in the
range expected at this redshift from hierarchical models of elliptical
galaxy evolution (Thomas 1999).  Ellipticals produce most of their
stars, and hence metals, in the first $\sim 0.5$~or 1 Gyr
(Rocca-Volmerange \& Fioc, 2000), and so at $z\sim 3$~they do not
resemble low redshift, extremely metal poor galaxies, but rather older
low-z starbursts.  Future LBG surveys with NIRSPEC will explore this
connection (Pettini et al. 2000b).  The ease with which our spectra
were obtained (1.3 hours of telescope time) demonstrates the great
potential for observation of typical ($K\simgt 20$) LBGs with NIRSPEC
or similar instruments.

\acknowledgements

It is a pleasure to acknowledge the hard work of past and present
members of the NIRSPEC instrument team at UCLA: Maryanne Angliongto,
Oddvar Bendiksen, George Brims, Leah Buchholz, John Canfield, Kim
Chin, Jonah Hare, Fred Lacayanga, Samuel B. Larson, Tim Liu, Nick
Magnone, Gunnar Skulason, Michael Specncer, Jason Weiss and Woon Wong.
In addition, we thank the Keck Director Fred Chaffee, CARA instrument
specialist Thomas A.  Bida, and all the CARA staff involved in the
commissioning and integration of NIRSPEC. We especially thank our
Observing Assistants Joel Aycock, Gary Puniwai, Charles Sorenson, Ron
Quick and Wayne Wack for their support.  
We thank M. Pettini, C.C. Steidel and
S.R. Heap for useful discussions.  HIT received support for this work
by the STIS Investigation Definition Team through the National Optical
Astronomical Observatories and by the Goddard Space Flight Center.

\references

Bechtold, J., Yee, H.K.C., Elston, R., \& Ellingson, E., 1997, ApJLetters, 477, L29

Calzetti, D.A., Kinney, A.L., \& Storchi-Bergmann, T.,
        1994, ApJ, 429, 482

Edmunds, M.G., \& Pagel, B.E.J., 1994, MNRAS, 21, 507

Ellingson, E., Yee, H.K.C,
        Bechtold, J., \& Elston, R., 1996, ApJLetters 466, 71; E96

Fitzpatrick, E.L., 1986, AJ, 92, 1068

Frayer, D.T., Papadopoulous, P.P., Bechtold, J., Seaquist, E.R., Yee, H.K.C., \& Scoville, N.Z., 
  1997, SJ, 113, 562

Kennicutt, R. 1983, ApJ 272, 54

Kobulnicky, H.A., Kennicutt. R.C. Jr., \& Pizagno, J.L., 1999, ApJ 514, 544; KKP99

Kurucz, R. L. 1993, CD-ROM 13, ATLAS9 Stellar Atmosphere 
 Programs and 2 km/s Grid (Cambridge: Smithsonian Astrophys. Obs.)

Leitherer, C., 1998, in Dwarf Galaxies and Cosmology, ed. Trinh X. Thuan, C. Balkowski, 
 V. Cayatte, \& Tranh Thanh Van (Paris: Editions Frontieres), in press (STScI preprint
 no. 1254)

Leitherer, C., Robert, C., \& Heckman, T.M., 1995, ApJS, 99, 173

Leitherer, C. et al., 1999, ApJS in press

Madau, P. Pozzetti, L., \& Dickinson, M., 1998, ApJ 498, 106

McGaugh, S., 1991, ApJ, 380, 140

McLean, I. S., et al., 1998, SPIE, Vol. 3354, 566 

McLean, I.S. et al., 2000a, PASP submitted

McLean, I.S. et al., 2000b, ApJLetters in press

Meurer, G.R., Heckman, T.M., Leitherer, C. Kinney, A., Robert, C., \& Garnett, D.R.,
  1995, AJ 110, 2665

Nakanishi, K., Ohta, K., Takeuchi, T.T., Akiyama, M, Yamada, T., \& Shioya, Y.,
  1997, PASJ, 49, 535

Olofsson, K. 1995, A\&AS, 111, 570

Osterbrock, D.E., 1989, Astrophysics of Gaseous Nebulae and Active Galactic
  Nuclei (Mill Valley: University Science Books)

Pagel, B.E.J., Edmunds, M.G., Blackwell, D.E., Chun, M.S. \& Smith, G.,
  1979, MNRAS, 189, 95

Pettini, M., Kellogg, M., Steidel, C.C., Dickinson, M., Adelberger, K.L., \& 
  Giavalisco, M., 1998, ApJ 508, 539; P98

Pettini, M., Steidel, C.C., Aldelberger, K.L., Dickinson, M., \& Giavalisco, M., 
  2000, ApJ, in press;  P2000

Pettini, M., et al., 2000b, in preparation

Rocca-Volmerange, B. \& Fioc, M., 2000, in "Toward a New Millenium in
Galaxy Morphology", Eds D.L. Block, I. Puerari, A. Stockton and D.
Ferreira (Kluwer, 2000)

Seitz, S., Saglia, R.P., Bender, R., Hopp, U., Belloni, P., \& Ziegler, B., 
  1998, MNRAS 298, 945

Steidel, C.C., Giavalisco, M., Pettini, M., Dickinson, M.,
        \& Adelberger, K.L., 1996, ApJ Letters 462, L17

Teplitz,H.I, Malkan,M.A., \& McLean,I.S., 1998, ApJ 506, 519 

Thomas, D., 1999, MNRAS 306, 655

Trager, S.C., Faber, S.M., Dressler, A., \& Oemler, A., 1997, ApJ 485, 92

Villa-Costas, M.B., \& Edmunds, M.G., 1992, MNRAS, 259, 121

Williams, L.L.R. \& Lweis, G.F., 1997, MNRAS 281, L35

Yee, H.K.C., Ellingson, E., Bechtold, R.G., Carlberg,R.G.,
        Cuillandre, J.-C., 1996, AJ 111, 1783

Yee, H.K.C., Ellingson, E., \& Carlberg,R.G.,
        1996, ApJS 102, 269

Zaritsky, D., Kennicutt, R.C., \& Huchra, J.P., 1994, ApJ, 420, 87

\clearpage

\begin{deluxetable}{lllrr}
\tablenum{1}
\tablecolumns{5}

\tablecaption{Emission Lines}
\tablehead{
\colhead{Line  (\AA)}&
\colhead{$\lambda_{obs}$~($\mu $m)}& 
\colhead{$z_{em}$}&
\colhead{$W_{rest}$\tablenotemark{a} (\AA)}&
\colhead{F\tablenotemark{b}}}

\label{tab:  fluxes}
\startdata

\oii    $\lambda 3726/3728$ &1.1.3898 &  2.72897 &  37$\pm 3$  &  12.74$\pm 1.22$ \nl
\hgamma~$\lambda 4340$    &  1.61828 &   2.72875 &   9$\pm 1$  &   1.61$\pm 0.17$ \nl
\hb~    $\lambda 4861$    &  1.81217 &   2.72774 &  26$\pm 4$  &   4.07$\pm 0.57$ \nl
\oiii~  $\lambda 4959$ \tablenotemark{c}   &  1.84913 &   2.72890 &  26$\pm 8$  &   4.01$\pm 1.30$ \nl
\oiii~  $\lambda 5007$    &  1.86678 &   2.72845 &  97$\pm 5$  &  14.73$\pm 0.78$ \nl
He I    $\lambda 5876$    &  2.19100 &   2.72873 &   3$\pm 1$  &   0.35$\pm 0.09$ \nl
[O~{\sc i}]~$\lambda 6300$ &  2.34949 &   2.72935 &  25$\pm 4$  &   3.06$\pm 0.46$ \nl
\nii~    $\lambda 6548$   &  2.44204 &   2.72944 &   7$\pm 2$  &   0.86$\pm 0.21$ \nl
\ha~    $\lambda 6563$    &  2.44750 &   2.72935 & 106$\pm 3$  &  12.56$\pm 0.37$ \nl
\nii~   $\lambda 6583$    &  2.45566 &   2.73031 &  10$\pm 2$  &   1.14$\pm 0.26$

\enddata

\tablenotetext{a}{Rest-frame equivalent width in \AA}
\tablenotetext{b}{Observed line flux in units of $10^{-16}$erg 
  s$^{-1}$ cm$^2$} 
\tablenotetext{c}{The blue wing of the 4959\AA~line
  is in a deep atmospheric absorption trough}

\end{deluxetable}

\clearpage

\figcaption[]{N4-band spectrum of MS1512-cB58.  20
  minutes of integration time.  Detected emission lines are indicated.
  Dotted lines show the locations of OH sky lines with residuals that
  are visible in the 2D spectrum.  The lower spectrum shows the
  1$\sigma$~errors, truncated at $14\times 10^{-17}$ergs cm$^{-2}$
  s$^{-1}$\AA$^{-1}$~for clarity, and arbitrarily shifted downward on
  the plot.  The increases in the errors occur at the position of the
  night sky lines and at positions of high atmospheric extinction.}

\figcaption[]{H-band spectrum of MS1512-cB58.  30
  minutes of integration time.  As in Figure 1, the 1$\sigma$~errors
  are truncated and shifted downward on the plot.
  The ``break'' in the continuum at 1.82\mic~is a residual from the
  atmospheric absorption correction, not a drop in the galaxy's SED.}

\figcaption[]{K-band spectrum of MS1512-cB58.  30 minutes of integration
  time.  As in Figure 1, the 1$\sigma$~errors are shifted downward on
  the plot.}

\clearpage

\begin{figure}[h]
\parbox{6in}{\epsfxsize=6in \epsfbox{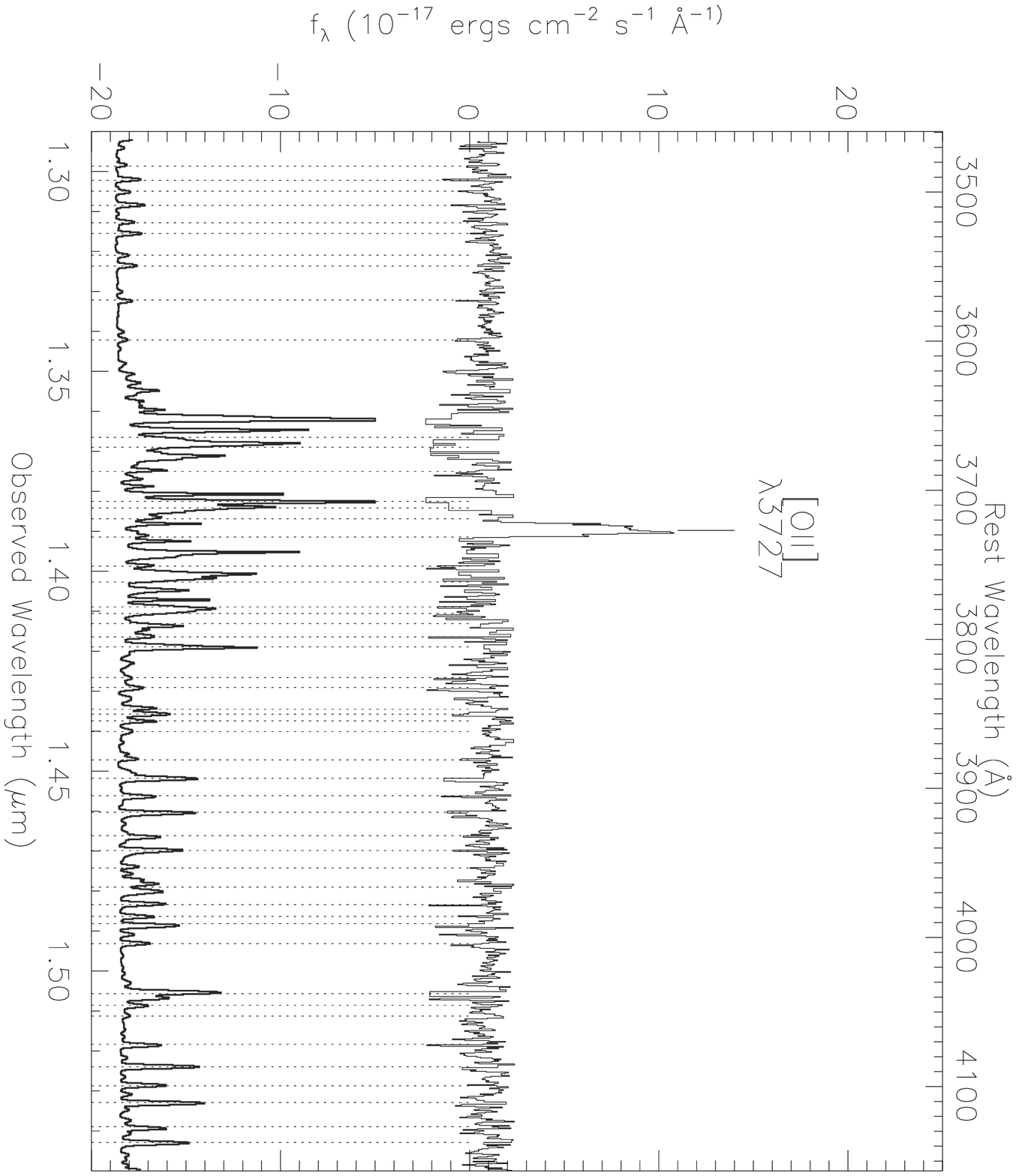}}
\label{fig:  j_spec}
\end{figure}

\clearpage

\begin{figure}[h]
\parbox{6in}{\epsfxsize=6in \epsfbox{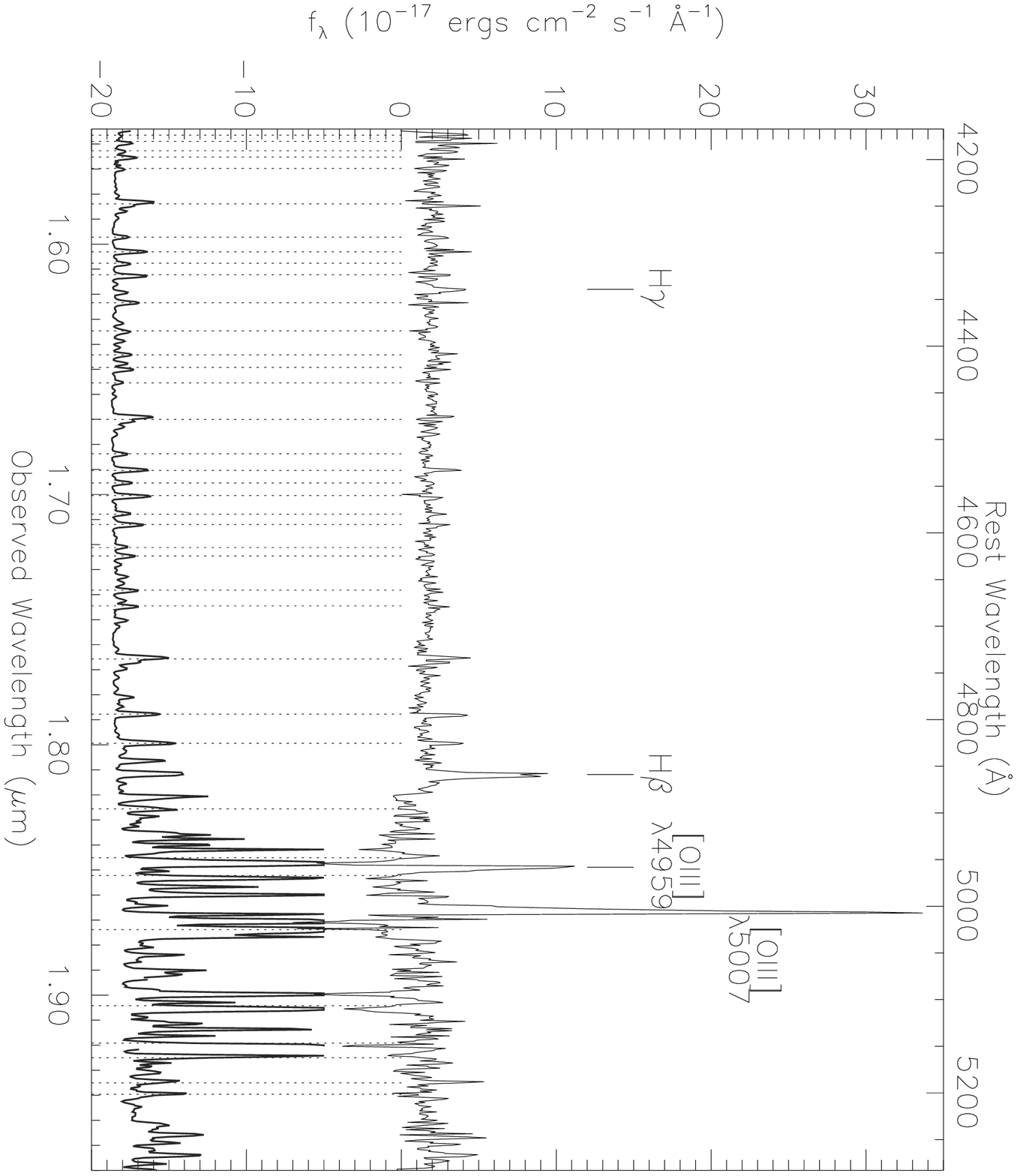}}
\label{fig:  h_spec}
\end{figure}

\clearpage

\begin{figure}[h]
\parbox{6in}{\epsfxsize=6in \epsfbox{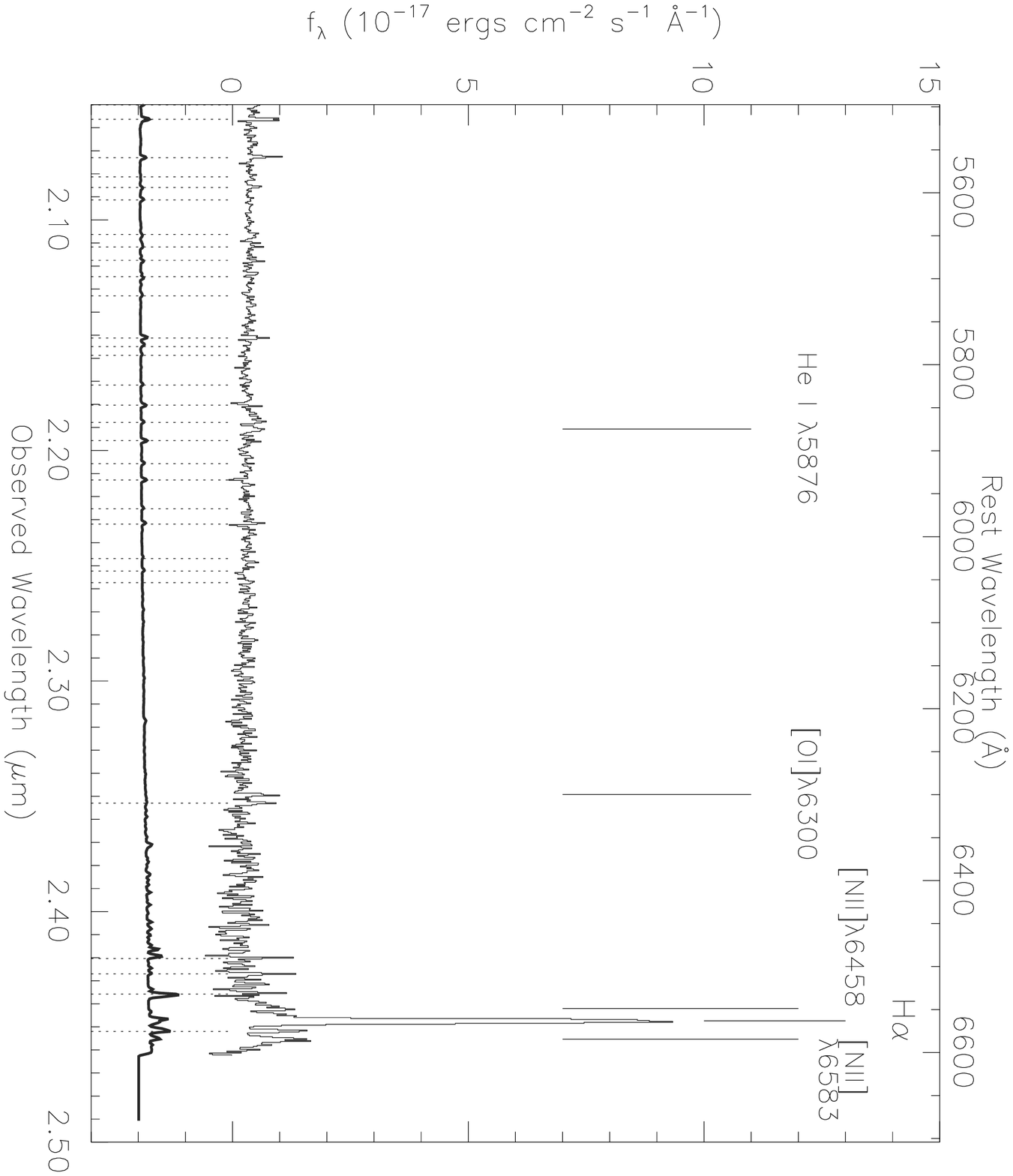}}
\label{fig:  k_spec}
\end{figure}

\end{document}